# A Study of Various Steganographic Techniques Used for Information Hiding

C.P.Sumathi[1], T.Santanam[2] and G.Umamaheswari[3]

[1,3]Department of Computer Science, SDNB Vaishnav College For Women, Chennai, India.
[2] Department of Computer Science, DG Vaishnav College For Men, Chennai, India.

## ABSTRACT

*The art of information hiding has received much attention in the recent years as security of information has become a big concern in this internet era. As sharing of sensitive information via a common communication channel has become inevitable, Steganography – the art and science of hiding information has gained much attention.  We are also surrounded by a world of secret communication, where people of all types are transmitting information as innocent as an encrypted credit card number to an online-store and as insidious as a terrorist plot to hijackers. Steganography derives from the Greek word steganos, meaning covered or secret, and graphy (writing or drawing) [1].  Steganography is a technology where modern data compression, information theory, spread spectrum, and cryptography technologies are brought together to satisfy the need for privacy on the Internet.  This paper is an attempt to analyse the various techniques used in steganography and to identify areas in which this technique can be applied,  so that the human race can be benefited at large.*

## 1. INTRODUCTION

The main purpose of Steganography, which means 'writing in hiding' is to hide data in a cover media so that others will not be able to notice it (Figure 1). While cryptography is about protecting the content of messages, steganography is about concealing their very existence [2]. The applications of information hiding systems mainly range over a broad area from military, intelligence agencies, online elections, internet banking, medical-imaging and so on. These variety of applications make steganography a hot topic for study. The cover medium is usually chosen keeping in mind the type and the size of the secret message and many different carrier file formats can be used. In the current situation digital images are the most popular carrier/cover files that can be used to transmit secret information.

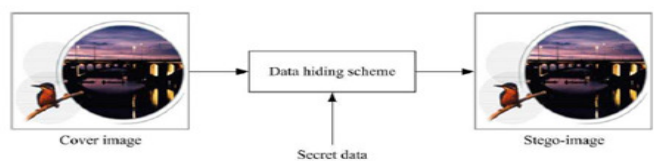

Figure1. Data Hiding Scheme (Intelligent Multimedia Data Hiding- Springer-ebook)

Steganography equation is 'Stego-medium = Cover medium + Secret message + Stego key'. The general model of data hiding can be described as follows. The embedded data is the message that one wishes to send secretly. It is usually hidden in an innocuous message referred to as a cover-text or cover-image or cover-audio as appropriate, producing the stego-text or other stego-object. A stego-key is used to control the hiding process so as to restrict detection and /or recovery of the embedded data to parties who know it [3].

Figure.2 depicts the different branches of steganography

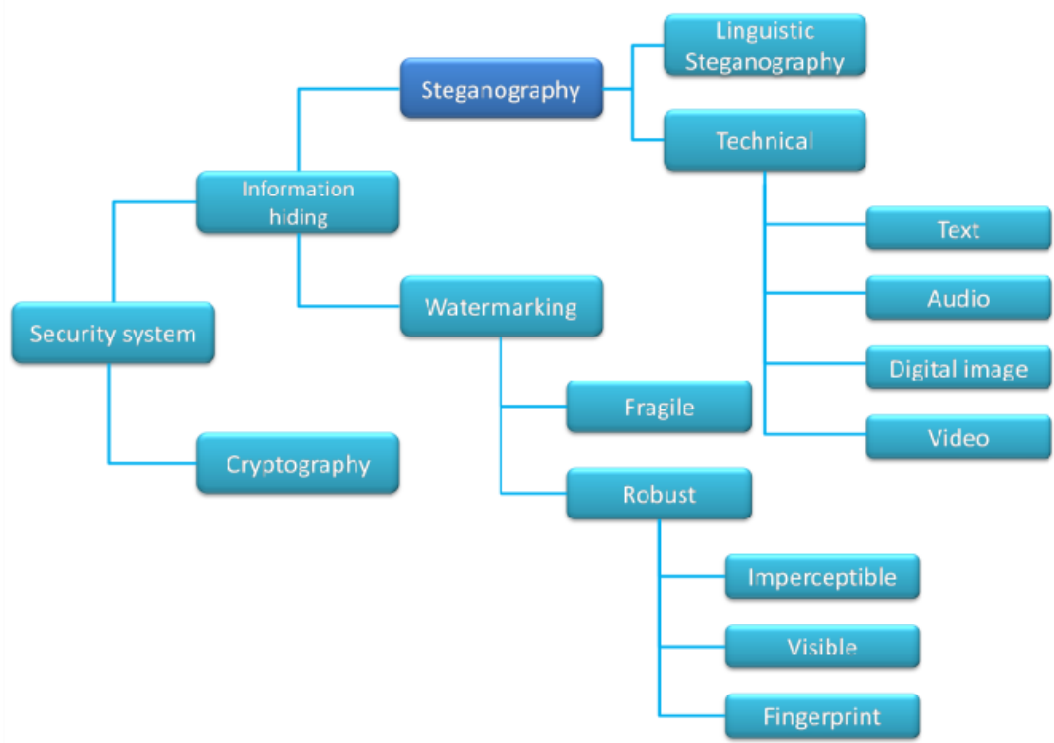

Figure 2: Steganography in security domain

(Source: www.ijarcsse.com Vol. 1, issue. 1 Dec 2011 )

While steganography can be achieved using any cover media, we are concerned with hiding data in digital images. The features expected of a stego-medium are imperceptibility and robustness, so that the secret message is known only to the intended receiver and also the stego-medium being able to withstand attacks from intruders. The amount of secret message embedded should be such that it doesn't reduce the quality of the stego image. This paper aims at studying the different techniques used in steganography for embedding data, their advantages and their differences. The goal of steganography is to embed secret data into a cover in such a way that no one apart from the sender and intended recipients even realizes there is secret data.

A few key properties that must be considered when creating a digital data hiding system are

- Imperceptibility: Imperceptibility is the property in which a person should be unable to distinguish the original and the stego-image.
- Embedding Capacity: Refers to the amount of secret information that can be embedded without degradation of the quality of the image.
- Robustness: Refers to the degree of difficulty required to destroy embedded information without destroying the cover image.

## 2. STEGANOGRAPHY TECHNIQUES

### 2.1. Classification of Steganographic Categories

Steganography is classified into 3 categories,

- Pure steganography where there is no stego key. It is based on the assumption that no other party is aware of the communication.
- Secret key steganography where the stego key is exchanged prior to communication. This is most susceptible to interception.
- Public key steganography where a public key and a private key is used for secure communication.

### 2.2 Classification of Steganographic Methods

Steganography methods can be classified mainly into six categories, although in some cases exact classification is not possible [2].

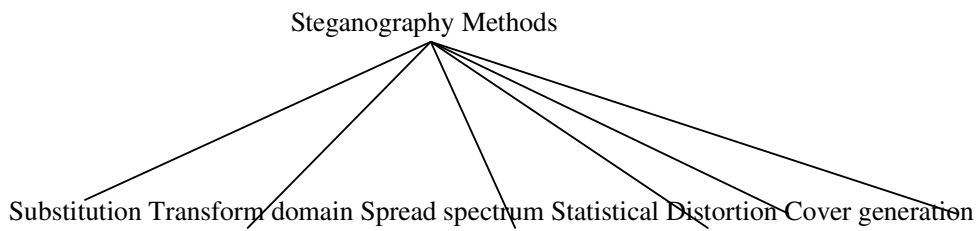

Steganography Methods

Substitution  Transform domain  Spread spectrum  Statistical  Distortion  Cover generation

- Substitution methods substitute redundant parts of a cover with a secret message (spatial domain).
- Transform domain techniques embed secret information in a transform space of the signal (frequency domain)
- Spread spectrum techniques adopt ideas from spread spectrum communication.
- Statistical methods encode information by changing several statistical properties of a cover and use hypothesis testing in the extraction process.
- Distortion techniques store information by signal distortion and measure the deviation from the original cover in the decoding step.

- Cover generation methods encode information in the way a cover for secret communication is created.

## 3. REVIEW

This paper analyses the various articles on steganography which help understand the topic in a new perspective.

### 3.1 Substitution Methods (Spatial-Domain)

In Mamta Juneja et. al's [4] research paper a secured robust approach of information security is proposed. It presents two component based LSB ( Least Significant Bit )methods for embedding secret data in the LSB's of blue components and partial green components of random pixel locations in the edges of images. An adaptive LSB based steganography is proposed for embedding data based on data available in MSB's of red, green, and blue components of randomly selected pixels across smooth areas. It is more robust as it is integrated with an Advanced Encryption Standard(AES).

In P.Thiyagarajan et.al's [5] scheme a new high capacity Steganographic scheme using 3D geometric models is proposed. The algorithm re-triangulates a part of a triangular mesh and embeds the secret information into newly added position of triangular meshes. This algorithm also resists against uniform affine transformations such as cropping, rotation and scaling. The stego key is generated from the message to be embedded. The vertices of the triangle are used for embedding.

In Shamim Ahmed Laskar et.al's[6] method data is embedded into the red plane of the image and the pixel is selected using a random number generator. It is almost impossible to notice the changes in the image. A stego key is used to seed the PRNG (Pseudo Random Number Generator) to select pixel locations. This paper focuses on increasing the security of the message and reducing distortion rate.

In S.Shanmuga Priya et. al's [7] article the authors propose a novel method based on LSB. Data embedding is performed using a pair of pixels as a unit, where LSB of the first pixel carries one bit of information and a function to two pixel values carries another bit of information. The proposed method shows better performance in terms of distortion and resistance against existing steganalysis. Embedding is done in the sharper edge regions using a threshold. PSNR value is compared for adaptive and non-adaptive techniques of data hiding in gray scale & color images.

In B.Sharmila et. al.'s [8] article, the authors propose an algorithm which works on colour images (JPEG). The edges are chosen for data hiding to improve robustness. The regions located at the sharper edges present more complicated statistical features and are highly dependent on the image contents. It is also more difficult to observe changes at the sharper edges than in smooth regions. In the embedding procedure, the RGB components are separated, and based on a shared key, one/more components are selected. The cover image is divided into non-overlapping blocks. Each block is rotated by a random degree determined by a secret key. The resulting image is rearranged as a row vector V by raster scanning. The secret message is encrypted and by using LSBMR, 2 secret bits can be embedded into each embedding unit. The message is embedded after calculating the capacity estimation using a threshold.

In Shweta Singhal et.al's [9] paper a new image steganography scheme is proposed in the spatial domain. In the technique, one byte of blue factor of pixels of an image have been replaced with secret bits of text data, which results in better image quality. A stego key is used for security purposes.

In Fahim Irfan Alam et. al's paper [10] the authors suggest noise filtering in the beginning before embedding. After extraction at receiving end, ARQ (Automatic Repeat Request) is used for error detection & correction. For secure transmission of data, encryption & data hiding are combined in a single step. Host image and secret data are converted into bit stream. Before encryption of secret data median filtering is used. The input values are converted to ASCII and then to binary, the host image RGB values are converted to binary. Substitution is performed character by character using encryption key. The LSB of every pixel octet is replaced by secret bit stream. Error detection and correction ensures correct transmission of data.

In Rajkumar Yadav et. al.'s paper [11], the authors present a study of a new method for insertion of message in an image. The last two bits of pixel value are used for insertion and retrieval of message. If the last two bits of pixel value are 00 or 10 , we can insert 0, else by adding /subtracting 1 at that pixel value we can insert 0. Similarly 1 is inserted if last two bits are 01 or 11. For increased security, message is embedded at pseudo random locations. The message is retrieved similarly based on the pixel values of the last two bits.

In M.B.Ould MEDENI et.al.'s article [12], the authors propose a novel method for hiding information within the spatial domain of the gray scale image. The Pixel Value Differencing (PVD) method segments the cover image into nonoverlapping blocks containing two connecting pixels and modifies the pixel difference in each block (pair) for data embedding. While embedding secret data, each pixel is split into two equal parts. The number of 1's in the most significant part is counted and the secret message is embedded in the least part according to the number of corresponding bits. The proposed method is based on four-pixel differencing and LSB substitution.

In Weiqi Luo et. al.'s paper [13], the authors propose an edge adaptive scheme which can select the embedding regions according to the size of the secret message and the difference between two consecutive pixels in the cover image. In the data embedding stage, the scheme first initializes some parameters, which are used for estimating the capacity of the selected regions. Finally stego image is obtained after pre-processing. A region adaptive scheme is applied to the spatial LSB domain and the difference between two adjacent pixels is used as a criterion for region selection and LSBMR (LSB Matching Revisited) as the data hiding algorithm.

In C.H.Yang et. al.'s article [14], a predictive method to enhance the histogram-based reversible data hiding approach is proposed. Two interleaving predictive stages are used. Most pixels are predicted by their two neighbourhood pixels and four neighbouring pixels in the column-based and chess-board based approach. The difference value of each pixel between the original image and the stego-image remains within ± 1. In interleaving predictions, pixels in odd columns will be predicted by pixels in even columns or vice versa. In the embedding process predictive error values of odd columns are used to generate a histogram to embed secret data. The predictive error values are converted to get the stego-image.

In Venkata Abhiram.M et. al.'s [15] article , the authors propose a randomization technique that makes use of RGB values of colour images to enhance imperceptibility. In the three channels RED, BLUE, GREEN the LSB of any one of the 3 channels is used as a pointer to decide embedding capacity in the other two channels. In the randomization technique, the LSB of any one of the channels (RGB) are used to indicate how data has to be hidden in the remaining 2 channels. If the last two bits of the channel are 00 there is no hidden data, if it is 01 data is embedded only in channel 2, if it is 10 data is embedded in channel 1 and if it is 11 data is embedded in both the channels. Three methodologies are used. They are, 1. RED is used as default pointer. 2. User selects any channel as pointer. 3. Pointers are chosen based on a cyclic sequence and data is embedded. Images were taken and same size data is embedded using all methodologies. Based on the histogram study and the values of MSE and PSNR (Mean Square Error and Peak Signal to Noise Ratio) the 3$^{rd}$ method i.e. the randomized method has better secrecy and performance with enhanced embedding capacities.

In G.Sahoo et. al.'s [16] article the authors recommend the use of a movie clip as carrier file to increase the capacity of secret data. The methodology works on the concept of replacement of entire non-sensitive pixel and the substitution of some part of the sensitive pixel with secret data. A movie clip is a temporal sequence of two dimensional samples of visual field with each sample being a frame of the movie. The parts of a movie clip can be divided into moving and static parts. The static and the dynamic parts can be obtained through Pixel Level Analysis, Likelihood Analysis or Colour Histogram Technique and stored in a static and dynamic buffer. In static portion embedding process one pixel is used to store three characters using the formula $x_{ij} = i+(j–1)*d$ where i is the initial location, j is character of the secret data and d is the distance between two embedding pixels. In dynamic portion embedding MSB method is used. A different stego-key is used for the dynamic portion. Main advantage of this method is more hiding capacity.

In Jasvinder Kaur et.al.'s article[17] the authors analyse different steganographic techniques based on digital logic and proposes a new enhanced steganographic technique based on it. The carrier image is selected depending on the information to carry. This technique uses digital operations based on logic gates and shift operators to embed/derive the hidden information from image data. Depending on the size of the information to embed the carrier image is divided into rows and data is embedded using digital operations.

In Hao-Tian et. al.'s paper [18] the authors suggest a method using adjacent bin mapping that is applied to 3D geometrical models. With the popularity of 3D scanning and modelling techniques steganography in 3D geometrics using adjacent bin mapping is analysed. Elements in two adjacent bins are mapped to each other for data embedding. Elements in the original object are bijectively mapped to those in the stego object. To apply the ABM method to digital images, in which the pixel values are represented by integers, the bin size $\Delta$ is set to 1 to minimize distortion. Since it is a kind of LSB method it is named LSB$^+$ algorithm. Given a gray scale image, its histogram is calculated by counting the pixels with the same value, i.e. the amount of pixels in each bin. The histogram of the cover image is preserved by restricting the amounts of 1's and 0's.

In Bawankar Chetan.D et. al.'s article [19] a prioritized sub blocks by pattern matching scheme is used to embed the code and a micro controller used for sake of security where it transmits a pre-programmed key at the beginning of each process. The steganalysis algorithm is continued only if the received key is correct at the destination; otherwise retrieving secure information from cover image is not accomplished. Perceptual multimedia sources enables data embedding as well as lossy compression either imperceptibly or with a controllable amount of perceptual degradation, whereas non-perceptual sources like text and executable codes require loss-less processing, transmission and storage. The image is divided into blocks. A block is selected for embedding based on a pattern. Several blocks are rejected due to various reasons like high visibility, non reversible at receiver side etc. In each sub block only the middle pixel is selected for hiding information. The secret message is encrypted to enhance security. In this paper the authors propose a steganographic model in which the hidden message can be composed and inserted in the cover in real-time. This is realized by designing and implementing a secret key steganographic micro architecture employing Field Programmable Gate Arrays FPGA. Four steganographic algorithms were found to be suitable for hardware implementation.

In Tanmay Bhattacharya et. al.'s article [20] the authors use a session based encryption and cross fold transposition for embedding. The secret text is converted to its binary form and cross fold transposition is performed. This binary form is perturbed by genetically generated session-key and embedded within the host image. For extraction both the stego image and the original image along with the session-key is sent.

In Chin-Chen Chang et.al.'s paper[21] an adaptive method is proposed. Data is hidden based on codeword grouping. A set of code words generated using palette generation algorithm is employed in index-based images. A code word grouping based steganographic scheme for index encoding images is presented. The relationship of code words is explored to group different member sub-clusters. The size of the sub-cluster determines the hiding capacity. To enhance hiding capacity sub-clusters with larger members are grouped together & sub-clusters with smaller members are grouped together. In the embedding procedure the sub-cluster to which the closest searched codeword belongs is identified, and the original encoded codeword is modified to hide secret message. The number of sub-cluster members indicates how many bits of secret message can be embedded. A set of thresholds is used to determine members of sub-cluster. Therefore choosing an adequate threshold is important. To improve security the sequence of embedding pixels is reorganised using a pseudo random generator.

In R.Chandramouli et.al.'s work [22] the authors analyse adaptive steganographic techniques which deal with hiding a message without causing statistically significant modifications. It is e-secure if the relative entropy of the probability distribution of cover objects and stego-objects is < or = e. LSB steganography is most simple and straight forward. The steganographic capacities of LSB based images are analysed using probability techniques & performance of steganalysis is also measured. Only framework of a passive warden is analysed. For security, message is encrypted before it is embedded. LSB based technique changes pixel value by ± 1 or leave them unchanged. The goal of a steganalyst is to estimate if I has hidden data. (I - Index set that denote the mean subtracted cover image)

Table 1. Substitution Based Steganographic Methods listed in chronological order starting from latest.

| S.No | Author | Year | Method used | Advantage | Other parameters |
|---|---|---|---|---|---|
| 1. | Mamta Juneja et.al.[4] | 2013 | Two component based LSB | More robust as algorithm is integrated with Advanced Encryption Standard (AES) | Ave. PSNR >52 |
| 2. | P.Thiyagarajan et.al [5] | 2013 | Scheme using 3D geometric models. | Resistance against uniform affine transformations such as cropping, rotation & scaling. | PSNR > 55 db  MSE 0 or between 0 & 0.1 |
| 3. | Shamim Ahmed Laskar et.al [6] | 2013 | Data embedding in the red plane of the image selected using PRNG | Increases security with reduced distortion rate | PSNR 58.8 db  MSE 0.0854 % |
| 4. | S.Shanmuga Priya et. al [7] | 2012 | Embedding done in the sharper edge regions using a threshold | Better performance in terms of distortion and resistance against existing steganalysis. | Non adaptive technique has more PSNR & less MSE than adaptive technique. |
| 5. | B.Sharmila et. al.[8] | 2012 | Edge regions selected for embedding using LSBMR ( LSB Matching Revisited) | Embedding capacity is more.  Improved security due to encryption. | MSE 0.013168  PSNR 66.96 dB  630 pixels embedded in a 256x256 image. |
| 6. | Shweta Singhal et.al [9] | 2011 | 1 byte of blue factor of pixels are replaced with secret bits. | Better image quality & secure. | PSNR 47.94 db  MSE 1.056 |
| **7.** | Fahim Irfan Alam et al[10] | 2011 | Noise filtering before embedding combined with encryption. | Error detection & Noise free transmission. | Success rate of >83% with different types of images. |
| 8. | Rajkumar Yadav et al.[11] | 2011 | A novel approach for image steganography In spatial domain using last two bits of pixel value | Limited changes in cover-image. | Chance of no change in pixel value is 50%  (7$^{th}$ & 8$^{th}$ bit) |
| 9. | M.B.Ould MEDENI et.al.[12] | 2010 | A novel steganographic method based on Pixel Value Differencing(PVD) | Imperceptibility as there is very less difference between original and stego image 40017 bits | 42.68 dB PSNR  40017 bits embedded in a 512x512 image. |
| 10. | Weiqi Luo et. al.[13] | 2010 | An Edge adaptive scheme for Region selection and LSBMR for data | Improved Visual Quality and security of secret | 54.1 PSNR  Embedding capacity 50% of |

| | | | embedding | message | cover image |
|---|---|---|---|---|---|
| 11. | C.-H.Yang et. al.[14] | 2010 | Improving histogram based reversible data hiding by interleaving predictions | Larger embedding capacity & better image quality. | 48.82 dB PSNR 99947bits embedding capacity (512x512 Image) |
| 12. | Venkata Abhiram.M et. al.[15] | . 2009 | Pixel Intensity based steganography with improved randomness. | Imperceptibility Embedding capacity is more. | 55dB app PSNR 0.1790 app MSE |
| 13. | G.Sahoo et. al.[16] | 2009 | Data embedded in static & dynamic portions after place analysis | Very high embedding capacity with security | Embedding position is calculated using $X_{ij}=i+(j-1)*d$ |
| 14. | Jasvinder Kaur et.al.[17] | 2009 | Embedding using digital operations are compared | More embedding capacity | Embedding capacity 1165084 bits .Image size 512 x 512 |
| 15. | Hao-Tian et. al.[18] | 2009 | Steganography in 3D geometrics & images using adjacent Bin Mapping ($LSB^+$ algorithm) | Statistics of image preserved | PSNR>60 dB Capacity 1200 bits (512x512 image) |
| 16. | Bawankar Chetan.D et. al.[19] | 2009 | Steganography Algorithm using Pattern Matching with External Hardware | External security using Hardware ¼ the size of the host image | 35.56dB 0.0013 MSE |
| 17. | Tanmay Bhattacharya et. al.[20] | 2009 | A hiding technique using bit level cross fold transposition and genetic algorithm | Embedding capacity & security of text is more compared to S-Tools. | Embedding capacity 108,242bytes Image size 732x486 |

| 18. | Chin-Chen Chang et.al.[21] | 2004 | Code word grouping –palette generation algorithm Encoded codeword is modified to hide secret message | Good Image Quality & Better Hiding Capacity | PSNR 36.34 Capacity 355588 bits Cover image:512x512 |
| --- | --- | --- | --- | --- | --- |
| 19. | R.Chandramouli et.al.[22] | 2001 | Adaptive Steganography | Imperceptibility | Analysis of pd & pf. probability of correct detection probability of false detection |

### 3.2. Transform Domain Methods

In Hemalatha.S et.al's [23] paper, the authors propose a method that uses two gray scale images of size 128 x 128 that are used as secret images and embedding is done in RGB and $YC_bC_r$ domains. The quality of stego images are good in RGB domain by comparing the PSNR values. The authors have used Integer Wavelet Transform (IWT) to hide secret images in the color cover image. The authors have compared the PSNR values and image quality when embedding is done in the RGB and $YC_bC_r$ domains.

In another article by Hemalatha .S et. al. [24] Integer Wavelet Transform (IWT)─have been suggested to hide multiple secret images and keys in a color cover image which is more efficient. The cover image is represented in the $YC_bC_r$ color space. Two keys are obtained, encrypted and hidden in the cover image using IWT.

In Keith.L. Haynes 's article [25] the author studies the use of image steganography to breach an organization's physical and cyber defences. The proposed method utilizes computer vision and machine learning techniques to produce messages that are undetectable and if intercepted cannot be decrypted without key compromise. To avoid detection DWT (Discrete Wavelet Transform) is used. The goal of a computer vision system is to allow machines to analyze an image and make a decision as to the content of that image. The computer vision can be categorized as Model-Based & Appearance Based which uses example images and machine learning techniques to identify significant areas or aspects of images that are important for discrimination of objects contained within the image. Machine learning is different from human knowledge/ learning. A computer has to make decision of the presence of a face based on the numbers contained in a 2D matrix. The feature is identified by using Haar feature selection. The goal is to identify the set of features that best distinguishes between images in the different classes. In the proposed method the cover image does not contain a secret message, rather the classification of the image yields the hidden message. Since the proposed algorithm utilizes ordinary unmodified images, there are no inherent indicators of covert communication taking place.

In S.Arivazhagan et. al.'s work [26] the authors propose a method that works in the transform domain and attempts to extract the secret almost as same as the embedded one, maintaining minimal changes to cover image by using techniques like median maintenance, offset & quantization. A modified approach for embedding colour images within colour images is proposed and it overcomes the limitations in embedding. Arnold Transform is applied on the secret image to increase robustness. This transformed image is then split into the three colour planes R, G, B and are subjected to DWT individually, converted to bit stream and then concatenated to be embedded in the cover image which is also subjected to DWT.

In Anindya Sarkar et. al.'s paper [27] the authors propose a Matrix Embedding with Repeat Accumulate (ME-RA) based steganography in which the host coefficients are minimally perturbed such that the transmitted bits fall in a coset of a linear code, with the syndrome conveying the hidden bits. The hiding blocks are pseudo-randomly chosen. A powerful repeat accumulate code is used for error correction. The authors have compared QIM (Quantization Index Modulation) and ME-RA methods. The comparisons with a slight modification of the ME-RA (puncture and non-shrinkage) methods with different decoding methods are also tabulated. The authors highlight that the use of ME instead of QIM within the YASS (Yet another Steganographic Scheme) that provides improved steganalysis performance but software complexity is more.

In Prosanta Gope et. al.'s article [28], the authors introduce an enhanced JPEG steganography along with a suitable encryption methodology using a symmetric key cryptographic algorithm. The JPEG cover image is broken into 8 x 8 blocks of pixel. DCT is applied to each block and quantization is done and data is encrypted using a new encryption method which uses CRC checking.

In Po-Chyi et.al.'s article [29] the authors compare the advantage of embedding in JPEG 2000 images with the previous approach of embedding in JPEG images. Most of the steganographic methods are based on JPEG because as a block DCT codec JPEG lends itself a good candidate for information hiding due to its fixed block structure. JPEG 2000 which is an upcoming still image coding standard can be used to hide high volume data. If information is embedded in the output of tier-2 coding, i.e. the JPEG 200 packets, it can be guaranteed that all the embedded information will be received without error and in correct order. But, difficulty lies in the modification of packets for embedding, since the bit-streams are compactly compressed by the arithmetic coder. Careless modification would result in failure of expanding compressed image. In the embedding process the image is decomposed using wavelet transform. (Number of wavelet decomposing levels & image size should be related to the host image), Lazy Mode Coding (Magnitude Refinement pass is suitable for steganographic purposes) is used for embedding.

In Hideki Noda et.al.'s paper [30] the authors propose a method that is based on a seamless integration of JPEG2000 lossy compression scheme and bit-plane complexity segmentation (BPCS) steganography. In bit-plane decomposition an n bit image is decomposed into a set of n binary images by bit slicing operations, combined with replacing binary data in LSB bit planes with secret data. The BPCS steganography uses bit-plane decomposition and characteristics of human vision. In JPEG 2000, wavelet coefficients of an image are quantized into a bit-plane structure. Each bit plane of the cover image is segmented into small size 8x8 blocks and are classified into informative / noise like blocks, using a threshold of the complexity $\alpha_{0 \text{ (e.g. value of } \alpha_0 \text{ 0.3 } \alpha_{max})}$ $\alpha_{max}$ is the possible complexity value. The secret file is segmented into a series of blocks containing 8 bytes of data that are regarded as 8x8 binary images. If secret block is less complex than the threshold $\alpha_0$, conjugate (XOR) it to make more complex. ($\alpha = \alpha_{max} - \alpha$). The image will now be a conjugated image. Replace each noise like block in the bit planes with a block of secret data. If block is conjugated store it in the conjugation map. Blocks can be randomly selected by using a random-number generator. Also embed the conjugation map with secret data (usually the first noise like block). Secret data is embedded after tier-2 encoding.

Table 2. Transform Domain Steganographic Methods in chronological order starting from latest

| S.No | Author | Year | Method used | Advantage | Other parameters |
|---|---|---|---|---|---|
| 1. | Hemalatha.S et.al [23] | 2013 | Integer Wavelet Transform is used to compare embedding in two different domains. | Quality of image is compared by embedding in RGB and $YC_bC_r$ domains. | PSNR in RGB 47<br><br>PSNR in $YC_bC_r$ 41 |
| 2. | Hemalatha .S et. al. [24] | 2013 | Multiple secret images are hidden in a cover image using IWT. | More efficient and secure as encryption is used. | Ave PSNR 44.7db |
| 3. | Keith.L. Haynes et. al.[25] | 2011 | Using Image Steganography to Establish Covert Communication Channels | Covert communication<br><br>Security by added encryption | Min. Image Size (128x128)<br><br>10 Haar features depicted<br><br>Final class recognition 11.5% |
| 4. | S.Arivazhagan et. al. [26] | 2011 | Colour Image Steganography Using Median Maintenance | Increased Security due to random selection<br><br>Minimal Changes in the cover image | PSNR 60.23 dB (app) |
| 5. | Anindya Sarkar et. al.[27] | 2010 | Matrix Embedding with Repeat accumulate codes | Robust against various global attacks | LLR computation (Log of 0 or 1 transmitted ) & bpnc (hiding rate) is compared with different decoding & hiding methods |
| 6. | Prosanta Gope et. al.[28] | 2010 | Enhanced JPEG steganography with suitable encryption | Added security using encryption | Bit Error Rate, MSE and PSNR values are computed |

| 7. | Po-Chyi et.al.[29] | 2003 | Steganography in JPEG 2000 compressed images | Controllable distortion<br><br>Can embed high volume data. | PSNR is 38.87 dB for 0.75 bit rate and increases for increasing bit rate.<br><br>Lazy mode coding |
| --- | --- | --- | --- | --- | --- |
| 8. | Hideki Noda et.al.[30] | 2002 | BPCS steganography with compressed images | Image Quality is Good | Embedding Rate 11.2% |

### 3.3. Statistical Methods

In Tomas Filler et. al.'s work [31], the authors propose a practical methodology for minimizing additive distortion in steganography with general embedding operation which is more flexible and easy. Syndrome-Trellis Codes (STC) are used to improve the security of the system. STC divides the samples into different bins (binning) which is a common tool used for solving many information-theoretic and also data-hiding problems. The proposed method can be used in both spatial & transform domain. A proper distortion function is chosen which makes statistical detection difficult. Once the steganographer specifies the distortion function, the proposed framework provides all tools for constructing practical embedding schemes. The distortion function or the embedding operations need not be shared with the recipient.

In Jessica Fridrich et.al.'s research paper [32] the authors propose a reversible embedding scheme for VQ-compressed images that is based on side matching and relocation. The new method achieves reversibility without using the location map. Even a tiny distortion of the original content is not applicable in some sensitive applications such as military, medical / fine art data. Therefore the value of reversible methods of steganography is increasing. VQ (Vector Quantization) is a popular compression technique because of its simple encoding and decoding procedures. To achieve better imperceptibility the codebook is partitioned into several clusters before embedding. The input needed will be a VQ compressed image, a stream of secret bits, a super codebook SC, clusters of the super codebook SC and multiple hit maps. The output will be a VQ stego image. Block X in the cover image will fall into one of the three following cases. If X is equal to the $i^{th}$ codeword of $G_o$, the embedding process is invoked. If X is equal to the $i^{th}$ codeword of $G_1$, no secret bit can be embedded and a compensation procedure is needed to avoid conflicting with case 1. If X does not belong to $G_0 \cup G_1$, no secret bit can be embedded and X is skipped. Secret bits can be embedded only in case 1.

In Chin-Chen Chang et.al.'s article [33] a new approach to wet paper codes using random linear codes of small co-dimension is used which improves embedding efficiency is proposed. To prevent from attack, the selection channel should not be publicly available even in any partial form. A possible remedy is to select it according to some side information that is in principle unavailable to the attacker (e.g.) random or that cannot be well estimated from the stego image. Steganography with non shared selection channels requires codes for memories with defective cells also called wet paper codes. This paper provides a new tool for steganography a coding method that empowers the steganographer with the ability to use arbitrary selection channels while substantially decreasing the number of embedding changes. The algorithm combines wet paper codes with matrix embedding arbitrary selection channels and improved embedding efficiency using random linear codes of small co-dimension.

In Zhicheng Ni et.al.'s article [34] the authors present a lossless data hiding which is robust against JPEG / JPEG 2000 compression. The image is split into 8 x 8 blocks and each block is split into two subsets (A, B). For each block the difference value α is calculated where α is the arithmetic average of differences of pixel pairs within the block. This α is selected as a robust quantity for embedding the information bit. Each bit of the secret message is associated with a group of pixels eg. A block in an image. The bit embedding strategy used is as follows, If α is located within a threshold & to embed bit 1, shift α to right/left beyond a threshold by adding/subtracting a fixed number from each pixel value within one subset. To embed 0, the block is intact. If α is located outside the threshold, always embed 1 thus shifting the value α away beyond a threshold. Then error correction code is applied.

Table 3 Statistical Steganographic Methods in chronological order starting from latest

| S.No | Author | Year | Method used | Advantage | Other parameters |
|---|---|---|---|---|---|
| 1. | Tomas Filler et. al.[31] | 2010 | Additive distortion function in Steganography using Syndrome Trellis codes | Statistically Undetectable. Can be implemented in spatial & transform domains. | Relative payload α compared with relative wetness τ from the no. of changed wet elements |
| 2. | Jessica Fridrich et.al.[32] | 2006 | Matrix embedding with wet paper codes | Embedding Efficiency with minimal changes to the cover | Embedding efficiency $e(p)=p/R_a(p)$  Efficiency increases for shorter messages reaches a maximum & decreases |
| 3. | Chin-Chen Chang et.al.[33] | 2006 | Reversible embedding scheme for VQ-compressed images based on side matching and re-location. Uses location map. | Suitable for VQ compressed images due to its high utilization of local block properties | PSNR 31.16 dB  Payload 8707 bits (128x128) |
| 4. | Zhicheng Ni et.al.[34] | 2004 | Lossless Data Hiding | No salt pepper noise  Can resist JPEG compression 1024/512 bits | >38 dB  1024/512 bits |

## 3.4. Distortion Methods

In M.B.Ould MEDENI et.al.'s article [35], the authors use error correcting codes in steganographic protocols. An optimal code is one that makes most of the maximum embeddable (MLE). The method referred to as matrix encoding requires the sender and recipient to agree in advance on a parity check matrix H. The cover medium is processed to extract a sequence of symbols v, which is modified into s to embed the message m, s is sometimes called the stego-data, and modifications on s are translated on the cover-medium to obtain the stego-medium. Relation between steganographic algorithms and error correcting codes are discussed.

In D.P.Gaikwad et. al.'s paper [36]the authors propose image restoration technique in steganography. The image is blurred before hiding the message image using special point spread function and randomly generated key. Sequential LSB embedding in the R plane is done in this project. The number of rows and columns of the message image is encrypted in the first row of the cover image. Before inserting, the original message image is blurred using the specific PSF (Point Spread Function). The parameters used for blurring with PSF are used as keys during de-blurring. The secret key values are sent through a secure channel (Tunnelling). The secret image is recovered using the two keys and a third key, which is randomly generated and depends on the content of the hiding message.

Table 4 Distortion Steganographic Methods in chronological order starting from latest

| S.No | Author | Year | Method used | Advantage | Other parameters |
| --- | --- | --- | --- | --- | --- |
| 1. | M.B.Ould MEDENI et.al.[35] | 2010 | Use of error-correcting codes in steganography | Improved embedding efficiency  Security | Performance evaluation by using D, average distortion, E embedding rate & R expected number of changes in cover. |
| 2. | D.P.Gaikwad et. al.[36] | 2010 | Image blurring with sequential LSB embedding | More secure and robust than simple LSB | 7.3064e-008 random key value for a png image of 24 KB |

## 4. CONCLUSION & FUTURE WORK

In this paper different steganographic articles were studied and were categorized into different techniques. As many new application areas are identified like internet banking, mobile communication security, cloud security etc., the insight into the steganographic principles will definitely guide us to identify new areas and to improve its applications in the already existing application areas also.

International Journal of Computer Science & Engineering Survey (IJCSES) Vol.4, No.6, December 2013

## ACKNOWLEDGEMENTS

We thank everyone who gave support and ideas in bringing out this paper.